\documentclass[aps,pre,twocolumn,amsmath,amssymb]{revtex4}
\usepackage{bm}
\usepackage{graphicx}

\newcommand{\be}{\begin{eqnarray}}
\newcommand{\ee}{\end{eqnarray}}

\newcommand{\ba}{\begin{array}}
\newcommand{\ea}{\end{array}}

\newcommand{\ket}[1]{|#1\rangle}
\newcommand{\bra}[1]{\langle #1|}

\begin{document}
\title{The relation between the quantum games, communication complexity problems and Bell inequalities}
\author{Marcin Paw{\l}owski}
\address{ Institute of Theoretical Physics and Astrophysics, Uniwersytet Gda\'{n}ski, PL-80-952, Gda\'{n}sk
\\
National Center for Quantum Informatics of Gda\'{n}sk}
\begin{abstract}
We study the relation between the quantum games, communication complexity problems and Bell inequalities. In particular we are interested in answering the question
whether for every element of one of these groups there is a corresponding element in the other two. We show that there are cases where there is no such
relation. Moreover, in the communication complexity problems for which there is no Bell inequality the advantage of the
quantum strategies over the classical ones is much higher.
\end{abstract}
\maketitle

\section{Introduction}
The topics of quantum games, communication complexity problems and Bell inequalities has been studied extensively. The reason for it is to find what are the
basic differences between the quantum world and the classical one. Since the same physical principles are used in all these field it is straightforward
to assume that there is no qualitative difference between them. Some work has already been done in order to prove so \cite{BZPZ}. The aim of this paper is to generalize
the proofs of existence of correspondence where it does exist and give the counterexamples where it does not. Before we start it is important to clarify
what we mean by a quantum game, since there are two very different categories of problems with this name attached. The first one are noncooperative games
where the players compete against themselves \cite{Eis}. We agree with van Enk's and Pike's \cite{vEnk} criticism of these type of games that the quantization of classical game leads to a case
where a new game is created. A game which could be also performed using classical means. Such games are useless when it comes to defining the
differences between classical and quantum worlds and therefore will not be dealt with in this paper. The cooperative games (also called nonlocal games)
on the other hand are games where teams of players compete \cite{QG1,QG2,QG3,QG4,QG5,CVG}. One of the teams has quantum resources the other does not.
The idea is to prove that there are some games, with the same sets of rules for both teams (that is they both get classical inputs and
are required to give classical answers) in which the quantum team has the advantage over the classical one. Form now on by the term quantum games we will
mean the cooperative ones.

\section{Communication Complexity Problems and Quantum Games}
The proof that for every communication complexity problem there is a quantum game and {\it vice versa} is trivial.
In a quantum game each of the players gets one of the components of the vector $\vec{i}$ representing input data. And they need
to send a component of a vector $\vec{o}$ representing the output to some fair referee who computes a
known function $f(\vec{o})$. If $f(\vec{o})=g(\vec{i},\lambda)$ where $g$ is another publicly known function and $\lambda$'s
represent some initial data not send to any of the players, then the players win. The players are not allowed to communicate.
In the case of communication complexity problems the situation pretty much looks the same. The only differences are that
it is one (or more) of the players that has to compute $f(\vec{o})$ and give it to the referee and that a very limited amount
of communication is allowed. These differences are not very important when it comes to applications and one can consider
a quantum game as a communication complexity problem with the maximum number of bits allowed to send equal 0.
This is quite useful since proving that for every Bell inequality there exist a quantum game
automatically proves that there also is a communication complexity problem. Also looking at the games from the communication
complexity point of view can be useful.
\

Instead of considering games that are tailored to give the quantum team an advantage (but from the gamer point of view
are rather boring) let us take a look real game that involves cooperation but limits communication and see if the quantum
team has the advantage also there. The example that suits the best is bridge. Before showing how to employ a communication
complexity problem in bridge let us look at one of them form another perspective. In the problem described in \cite{BCD} A and B
are given two boolean variables $x$ and $y$ each and they both must compute the function $f=x_A\oplus x_B \oplus (y_A y_B)$.
If we would modify their task so that only B is giving the answer then $x_B$ has no influence on the complexity of the problem
and both quantum and classical teams have the same probability of success computing $f_m=x_A \oplus (y_A y_B)$ with one
way communication as they would have with $f$ and two way communication. The modified task can be seen as follows: if $y_B=0$
then B is interested in receiving form A the value of a function $g_0=x_A$; if $y_B=1$ then he is interested in
$g_1=x_A \oplus y_A$. Since A does not know $y_B$ she sends the important function in 50\% of the cases in the other cases
B can only guess the answer which gives him the overall success probability of 75\% which is less than the 85.4\% achieved
with quantum resources. The quantum protocol can be seen as encoding both functions $g_0$ and $g_1$ in one bit but allowing
B to decode the value of only one (and not always correctly). The knowledge of B which function is for him interesting is
the thing that gives him the advantage over his classical counterpart despite the fact that he sometimes gets the wrong answer.
\

Now we are going to get more technical and the readers
who are not bridge players are advised to skip the rest of this paragraph. During the auction phase the goal of each team is
to compute the contract that gives them the highest score. It can be viewed as computing the function
$C(h_1,h_2,h_3,h_4)$ where $h$'s denote players' hands. The amount of communication is limited by the rules of the auction
itself. Let us consider an example where it is the clearest. After a few calls it had become clear that N and S will play
$5\diamondsuit$ or small slam  $6\diamondsuit$. N the declarer has the stronger hand so it is up to him to decide which
contract to choose. He can use the Blackwood convention and call artificial 4NT to ask for S's aces in order to establish
the number of tricks they are going to take. In this particular case S cannot answer in a standard way since one may think
of situation where the call $5\heartsuit$ gives N the information that he cannot play higher than $5\diamondsuit$ which
is then impossible. Actually S can send to N only one bit of information (calling $5\clubsuit$ or $5\diamondsuit$).
Let us assume that information about S's hand send to N so far tells him that S has 7-8 points, 0 of them in $\heartsuit$s
and $\spadesuit$s and 4+ in $\diamondsuit$s, he also has 7 $\diamondsuit$s. He can only have aces in  $\clubsuit$s or $\diamondsuit$s
and the knowledge which aces are in his hand is crucial to N. Let us now prove that there are cases in which it is optimal
for N and S to use the modified version of the protocol from \cite{BCD} presented in the previous paragraph. Let $x_A$ be 1 if
S has $A\diamondsuit$ and 0 otherwise, and $y_A$ be 1 if S has $A\clubsuit$ and 0 otherwise. The value of $y_B$ is computed
by N from the cards in his hand. Example of hand leading to $y_B=0$ is
$\diamondsuit: Q,9,8,4,3,\ \heartsuit:K,Q,J,10,\ \spadesuit:A,K,Q,J$. N knows that E and W have only one $\diamondsuit$ (A or K)
if this is A than N and S loose one trick in $\diamondsuit$s and one in $\heartsuit$s and can bid no higher then $5\diamondsuit$.
If it is S who has the $A\diamondsuit$ then they can bid $6\diamondsuit$. Example of hand leading to $y_B=1$ is
$\clubsuit: K,Q,\ \diamondsuit: 10,8,4,\ \heartsuit:A,K,Q,\ \spadesuit:A,K,Q,J$. If S has both aces then there is a possibility
of one of adversaries having $\diamondsuit: K,Q,J$ and taking two tricks. If he has no aces then these aces in adversaries hands
will also take two tricks. If he has only one ace we need to consider two cases. First, he has $A\clubsuit$, but he has also
4+ points in $\diamondsuit$ so he has $K\diamondsuit$ and $J\diamondsuit$ which should allow N not to loose more than one
trick. Second, he has $A\diamondsuit$ and at least $Q\diamondsuit$ which should allow N not to loose any tricks in $\diamondsuit$s
and only one in $\clubsuit$s. As we see N is really interested in $x_A\oplus y_A$. If N and S share a system in a singlet
state their probability of choosing the optimal contract is higher than without it, by the power of the arguments form \cite{BCD}.
This example is not perfect since showing N the information he has before he calls 4NT can be very hard in the first part of the auction and
the probability of one of the adversaries having all three $\diamondsuit$s in the second case is quite low but it does show the point.
Moreover, this advantage is only present if
variables can have both values with the probabilities similar enough, so the players need to have some strategy
that basing on calls of all 4 players allows them to calculate the probability distribution of these variables and decide
whether it is better to use quantum protocol or not. The proof that such strategy exists and that there are cases when it tells
players to use quantum protocols is currently being completed (along with an example that suits better) an will be published elsewhere.
\

Why do we spend so much time on discussing the application of communication complexity problems in bridge? The reason is that
this game is a serious one and unlike all the other games where the quantum team has the advantage it is really being played. If
any pair would use this Quantum Bridge Convention (and it is possible with current state-of-the-art equipment) then it would
be a great promotion of Quantum Information Theory among the general public. Even the prohibition by World Bridge Federation of
the use of quantum resources would be interesting since it would be the first time that the quantum phenomena would be
considered in some everyday law.

\section{From Bell inequalities to Quantum Games}
It has already been proven that Bell inequalities of a certain type lead to communication complexity problems and therefore
quantum games \cite{BZPZ}. It is easy to generalize this results to the inequalities basing on correlation functions of the type
studied by Sliwa \cite{Sliwa} or Uffink \cite{Uffink}.
\

The type of inequalities studied by Sliwa \cite{Sliwa} is of the form
\be \label{Sliwa}
\sum_{x_1,x_2,...,x_n} g_{x_1,x_2,...,x_n} E(x_1,x_2,...,x_n) \leq B
\ee
where $x_j$ denotes the setting at $j$-th lab. Some of the correlation functions in these inequalities depend on less variables
but we can always add to the set of the possible values of $x$'s one more, meaning that the lab does not perform any measurement and returns the
outcome 1. In this case we can rewrite all the inequalities in the form (\ref{Sliwa}) and proceed in way described in \cite{BZPZ} to get
communication complexity problems. The number of the possible values of the $x$'s is irrelevant.
\

The case of quadratic inequalities based correlation functions studied by Uffink \cite{Uffink} is a little more complex. They can be
written down as
\be\nonumber
\left(\sum_{x_1,x_2,...,x_n} g_{x_1,x_2,...,x_n} E(x_1,x_2,...,x_n)\right)^2+\\ \label{Uffink}
\left(\sum_{x_1,x_2,...,x_n} h_{x_1,x_2,...,x_n} E(x_1,x_2,...,x_n)\right)^2\leq B
\ee
Fortunately for every set of $x$'s $|g_{x_1,x_2,...,x_n}|=0,1$ and $|h_{x_1,x_2,...,x_n}|=0,1$. Moreover
$\sum_{x_1,x_2,...,x_n} |g_{x_1,x_2,...,x_n}|=\sum_{x_1,x_2,...,x_n} |h_{x_1,x_2,...,x_n}|=N$. This allows us to
rewrite (\ref{Uffink}) as
\be \label{circ}
\left(P_g-\frac{1}{2}\right)^2+\left(P_h-\frac{1}{2}\right)^2 \leq \frac{B}{4N^2}
\ee
where $P_g$ denotes the probability of computing the function
\be
f_g=y_1y_2...y_nS[g_{x_1,x_2,...,x_n}]
\ee
where the probability distribution of $y$'s is uniform and the probability distribution of $x$'s is given by
\be
Q_q{x_1,x_2,...,x_n}=\frac{|g_{x_1,x_2,...,x_n}|}{N}
\ee
$P_h$ is defined analogically. Equation (\ref{circ}) defines a circle.
If some quantum state violates (\ref{Uffink}) then to this state there corresponds a point $P=(p_1,p_2)$ outside the circle.
The communication complexity problem corresponding to this inequality would be of the type discussed in \cite{BZPZ} but enriched
by one more variable $z$ being send to each party. Its value would tell the participants which function $f_g$ or $f_h$ to
compute. The probability of $z$ having one of the values (choice is arbitrary due to the symmetry of the circle) is
$\frac{p_1}{p_1+p_2}$.
\

For arbitrary Bell inequalities it is more straightforward to find quantum games than communication complexity problems.
Let us start from linear inequalities. Every linear Bell inequality can be written down in a form
\be \label{lin}
\sum_{\vec{s},\vec{r}}a_{\vec{s},\vec{r}}P(\vec{r}|\vec{s})\leq B
\ee
Vector $\vec{s}$ represents the settings at all the labs and $\vec{r}$ the results. The sum can be substituted by an
integral without any change in the arguments that follow. By substituting some probabilities
with identity minus the probability of complementary events one can always write (\ref{lin}) in such a way that all
$a$'s are nonnegative and $\sum_{\vec{s},\vec{r}}a_{\vec{s},\vec{r}}=1$. Therefore $a$'s can be interpreted as probabilities.
The quantum game can be constructed from this inequality in the following way. The source generates two vectors
$\vec{s}$ and $\vec{r}$ with a probability $a_{\vec{s},\vec{r}}$. One component of vector $\vec{s}$ is send to each of the
players. Now the source asks the players to guess the vector $\vec{r}$ in such a way that every player gives one of its
components. It is easy to see that the probability of success in such a game is LHS of (\ref{lin}) and the quantum
strategy is simply to measure a state that violates (\ref{lin}) in the basis given by $\vec{s}$ and return to the source
the measurement results.
\

Now let us move to the general form of Bell inequality. For simplicity we will represent the set of two vectors $\{\vec{r},\vec{s}\}$
by a single index $i$. The linear Bell inequality (\ref{lin}) looks now
\be
\sum_i a_i P(i)\leq B
\ee
and the general one
\be \label{gen}
F(P(0),P(1),....)\leq B
\ee
We conjecture that without the loss of generality we can write down any Bell inequality in the form (\ref{gen}) and that
the function $F$ can be expanded into
\be
F(P(0),P(1),....)=\sum_{k_0,k_1,....}c_{k_0,k_1,....}\prod_iP(i)^{k_i}
\ee
Then the general Bell inequality takes the form
\be \label{gen2}
\sum_{k_0,k_1,....}c_{k_0,k_1,....}\prod_iP(i)^{k_i}\leq B
\ee
By the power of the same arguments that were used while considering equation (\ref{lin}) we may have nonnegative $c$'s
satisfying $\sum_{k_0,k_1,....}c_{k_0,k_1,....}=1$ and having the interpretation of probabilities. The part
$\prod_iP(i)^{k_i}$ of the LHS of (\ref{gen2}) is the probability of success in $K=\sum_ik_i$ consecutive trails, where the
trail $i$ with the success probability $P(i)$ is repeated $k_i$ times. In order to construct the quantum game the source
sends $K$ vectors $\vec{s_n}$ (n=1,..K) with probability the
\be
P_K=\sum_{k_0,k_1,....}\delta(\sum_ik_i,K)c_{k_0,k_1,....}
\ee
where $\delta$ is a standard Kronecker delta. If a vector $\vec{s_n}$ corresponds to the index $i$ then it will be send $k_i$
times. Now the quantum game consists simply of $K$ runs of the game for linear Bell equations.
\

To sum this section up, we have proven that for any Bell inequality there exists a communication complexity protocol and
a quantum game and only the states that violate this inequality give the quantum teams advantage over the classical ones.

\section{From Quantum Games to Bell inequalities}

We start by giving an example of a Bell inequality derived from a quantum game. The example is interesting because it
is the first Bell inequality derived in such a fashion and that the quantum game it originates from is a continuous variable
game with the highest difference of success probabilities for quantum and classical case known so far.
\

Alice and Bob are each given a real number $x_A,x_B\in (0,2\pi]$. These numbers can represent points on a circle.
Alice is allowed to send only one bit of information to Bob who must then give the answer to the question whether the point
he has been given is closer to the Alice's point or to the point opposite to hers. The probability distribution of Alice's
point on the circle is uniform. Bob's point is located not farther than an angle $\eta$ form either Alice's point or from the
point opposite to hers. In this part of the circle the probability distribution for Bob's point is also uniform. The joint
probability distribution is described by the equation (\ref{joint})
\

The most general strategy in the classical case is to divide the circle in two parts. Then the bit that Alice sends to
Bob tells him in which part Alice is. Bob answers that his point is closer to Alice's than to the opposite one
if his point is in the same part of the circle. One can integrate over every possible combination of numbers that the parties can get and
find that the optimal choice is to divide the circle into two connected sets of the same size. In that case the probability of success is
\be\label{pc}
P_C=1-\frac{\eta}{2\pi}
\ee
The quantum strategy is for parties to share a singlet state and make measurements of this state in such bases that the distance
from the vector corresponding to the outcome 1 to some set point on the Bloch sphere is equal to the number given. Parties have
agreed before on the direction in which this distance is measured and on the plane that both the are going to take place in.
If Alice sends Bob her measurement outcome and Bob
answers that his point is closer to Alice's than to the opposite one if his measurement had yielded the same outcome.
One finds that the probability of success in that case is
\be
P_Q=\frac{1}{2}+\frac{\sin(\eta)}{2\eta}
\ee
The difference between these two probabilities is
\be
\Delta p=P_Q-P_C=\frac{1}{2}\left(\frac{\sin(\eta)}{\eta}+\frac{\eta}{\pi}-1   \right)
\ee
It reaches the maximum of $\Delta p\approx 8.02\%$ for $\eta\approx1.07$. The $\Delta p$ is by 17.4\% better than the one
from \cite{CVG}.
\

It is straightforward to get a Bell inequality from this type of games
\be \label{newbell}
\int dx_A dx_B \rho(x_A,x_B)P(f(x_B,e_A)=g(x_B,x_A))\leq P_C
\ee
where $e_A$ is what Alice sends to Bob, $g$ is the function they need to compute, $f$ is Bob's answer and $\rho(x_A,x_B)$ the
probability density of having input data $x_A$ and $x_B$. In the case of the game presented we have
\be \nonumber
\rho(x_A,x_B)&=&\frac{H(\eta-|x_A-x_B|)+H(\eta-|x_A+\pi-x_B|)}{8\pi\eta}
\\ \label{joint}
\\
g(x_A,x_B)&=&H(\eta-|x_A-x_B|)
\\
f(x_A,x_B)&=&\delta{e_A,e_B}
\ee
where $H$ is the Heavyside function, $e$'s denote the measurement outcomes (in the classical case the measurement is simply
looking in which part of the circle the point is), and $P_C$ is given by (\ref{pc}).
\

Now to answer the final question of this section: Is there a Bell inequality for every quantum game? The surprising answer is: No.
Consider a game that we will nickname: Quantum TV Show. Two players A and B have never met each other before. Alice will get
two bits $y_a=0,1$ and $x_a=0,1$. Bob will get only one bit $y_b=0,1$. Then Alice can send only one bit to Bob who has to give
the correct value of the function $f=x_a \oplus(y_a y_b)$. If he gets it right they both get a prize.
Before they start they can discuss their strategy over a public channel, but C (the Celebrity that runs the show) is listening.
When the strategy of A and B is ready the game starts. C sends them the bits prepared in such a way that their probability of success is
the lowest (the TV company does not want to spend their money on the prizes). In this game it is not the players' strategy
that is tailored according to the probability distribution of the input data but exactly the opposite. For this game
instead of considering two standard cases: the classical and the quantum one we will take into consideration much more
possibilities of resources available to players or C in order to find what traits of physical theories are useful for
communication complexity problems.

\subsection{Deterministic (D)}
If the world is deterministic and local (Newton's physics apply),
there are only two ''sensible'' functions $g_i$ of Alice's bits that she can communicate to Bob
\be
g_0=x_a \quad g_1=x_a \oplus y_a
\ee
Each of them is equal to $f$ for only one value of $y_b$ (different for both functions). If Alice sends function $g_i$ then Bob has all the information he
needs iff $i\oplus y_b=0$. If $i\oplus y_b=1$ the Bob needs to know the value $y_a$ to give the right answer. So the Celebrity knowing the strategy
(that includes the value $i$) chooses $y_b$ such that $i\oplus x_b=1$. Now Bob can only guess the value of $f$ but since the situation is purely deterministic
C knows exactly what Bob is going to say in that case and chooses $x_a$ and $y_a$ in such a way that Bob is always wrong. So the probability of success is
$P_D=0$.

\subsection{Random (R)}
If Alice and Bob can have access to the random/pseudorandom number generators the only thing that changes is that Bob can choose the value of $f$ that he
gives as an answer randomly, giving them success probability of $P_R=\frac{1}{2}$.

\subsection{Quantum (Q)}
If Alice and Bob have quantum resources then their situation improves. They can use the modification of the
protocol proposed in \cite{BCD} discussed earlier. Note that the probability of success in quantum case does not depend on the input data. If A and B share the state
\be
\rho_p=\frac{(1-p)}{4}\mathbb{I}+p\ket{\psi^-}\bra{\psi^-}
\ee
one can use the method described in \cite{BZPZ} to get the probability of success equal
\be
P_Q(p)=\frac{1}{2}\bigg(1+\frac{\sqrt{2}p}{2} \bigg)
\ee
since $0\leq p \leq 1$ one sees immediately that
\be
\frac{1}{2}\leq P_Q(p) \leq \frac{2+\sqrt{2}}{4} \approx 85.4 \%
\ee
so even the smallest $p>0$ gives an advantage over classical (D and R) protocols. One does not need $\rho_p$ to violate Bell's inequality or even
to be entangled. The maximal success probability for separable states is $P_{QS}=P_Q(\frac{1}{3}) \approx 61.8 \%$ for entangled states that do not
violate Bell's inequality $P_{QE}=P_Q(\frac{1}{\sqrt{2}}) = \frac{3}{4}$ and for states that do violate CHSH $P_{QV}=P_Q(1) \approx 85.4 \%$.

\subsection{Cryptographic (C)}
If we assume that during the strategy discussion phase A and B can use some cryptographic (classical or quantum) protocol that is unbreakable for C then
this case is equivalent to Alice and Bob sharing the same string of random numbers which are unknown to C. This is in turn equivalent to A\&B's
strategy being unknown to C. This reduces the problem to the classical case discussed in both \cite{BCD} and \cite{BZPZ} and yields the probability $P_C=\frac{3}{4}$.

\subsection{Nonlocal Hidden Variables (N)}
Let us assume that A and B share the state $\rho_p$ but C knows Nonlocal Hidden Variables (NHV) relevant to this state and local pseudorandom
number generators at A and B's. The world is deterministic but nonlocal. Though C knows the state of the world he cannot influence it (similarly to
case D). Let $B$ be the Bell operator for CHSH inequality
\be
B=A_0 B_0+A_0 B_1 +A_1 B_0 - A_1 B_1
\ee
then
\be \label{war}
\textrm{Tr}(B \rho_p)=2\sqrt{2}p
\ee
The probability of success for Alice and Bob if they choose the measurements $A_{y_a} B_{y_b}$ is equal to
\be
P_N(p)=\frac{\textrm{Tr}(A_{y_a} B_{y_b} \rho_p) \pm 1}{2}
\ee
where minus corresponds to the measurement $A_1 B_1$ and plus to the rest of the measurements. C will choose such $y_a$ and $y_b$ for which
$P_N(p)$ is minimal for the current NHV's. Let us assume that it is for $y_a=y_b=0$. We have that
\be \nonumber
\textrm{Tr}(A_{0} B_{0} \rho_p)&=&
\\\nonumber
\textrm{Tr}(B \rho_p)-
\Big(\textrm{Tr}(A_{0} B_{1} \rho_p)+\textrm{Tr}(A_{1} B_{0} \rho_p)-\textrm{Tr}(A_{1} B_{1} \rho_p)\Big)
&\geq&
\\
\textrm{Tr}(B \rho_p)-3=2\sqrt{2}p-3
\ee
The success probability is then
\be
P_N(p)\geq \sqrt{2}p-1
\ee
Now it is grater than zero (that is better than local deterministic case) iff $p \geq \frac{1}{\sqrt{2}}$, so only for states that violate CHSH.
For maximally entangled states it becomes $P_N(1)\geq \sqrt{2}-1\approx 41.4 \%$. The ''$\geq$'' instead of ''$=$'' comes form our lack of knowledge
about the actual structure of the NHV theory. So we assume the worst possibility
\be
P_N(p)=\max \{ \sqrt{2}p-1,0\}
\ee

\subsection{Conclusions}

The Quantum TV Show has been created basing on a communication complexity problem \cite{BCD} for which the quantum strategy provides the advantage
if the communicating parties share states that violate CHSH inequality. But after simple modifications we end up with a game
for which there are states that do not violate any Bell inequality and still give the Quantum team (Q) the advantage over
the classical teams (D) and (R). In this sense we conclude that there is no Bell inequality corresponding to this game. This is due to the
fact that here the source (Celebrity) depends on the measurement choices (the players strategy). This is well acceptable
in the case of quantum games but not in the case of Bell inequalities. The same procedure can be repeated for every pair
of Bell inequality and communication complexity problem leading to a new problem with no corresponding inequality. In this
new problem the quantum team will have a greater advantage over the classical one than in the original problem.
\

By studying the dependance of success probabilities on the resources we can establish the traits of a physical theory
that are useful in quantum games or communication complexity problems. We see that pure randomness gives the success
probability of 50\% while the correlated randomness (which is what the Cryptographic (C) case boils down to) increases
this to 75\%. The non-locality itself gives us only 41.4 \% (N), but when aided with randomness it becomes 85.4\% (Q).
\

Let us treat the Quantum TV Show game as a communication complexity problem.
The application of problems with sources dependant on the strategy can be found in the cases where we want to slow down
the calculations as much as possible. Nowadays the classical cryptography is a good example. From a public key it is
possible to calculate the private one, but it takes a lot of time and the designers of the cryptographic systems try to
stretch this time as much as possible. Note that if the two parties are going to discuss a strategy of computation after
they receive the data then it automatically increases the communication complexity so the assumption of the strategy
being set before the data is received is sensible. This strategy should in principle be available to the creators of the
cryptographic system that these parties are going later to crack. By knowing this strategy the system designer can choose
such system parameters that it will have exactly the same effect on cracking the system as the Celebrity's choice had on
the winning probability of the players. In this scenario quantum strategies give much greater advantage than with the source
independent of the choice of strategy. Furthermore, the advantage is always present (if the amount of noise is lower than 100\%)
so the construction of quantum devices giving the advantage is much simpler.

\section{Summary}
The aim of this paper was to study the relation between the quantum games, communication complexity problems and Bell inequalities.
In the process the following things have been shown:
\begin{itemize}
\item There is no significant difference between the quantum games and communication complexity problems.
\item Quantum resources can be used to gain an advantage in already existing games (bridge). This can be of a great value
for the popularization of Quantum Information Theory.
\item From every Bell inequality one can derive a quantum game and a communication complexity problem.
\item There are quantum games and communication complexity problems that do not correspond to any Bell inequality.
\item There is a new Bell inequality (\ref{newbell}) leading to a continuous variable quantum game with the highest
known difference of success probabilities between the quantum and classical scenarios.
\item The traits of physical theory that positively influence the success probability in communication complexity problems
are randomness, correlated randomness and nonlocality.
\item When the input data depends on the chosen strategy the advantage given by the quantum resources is greater than when
it does not.
\end{itemize}

The author wishes to thank \v{C}. Brukner, M. Kurant, T. Paterek and M. \.{Z}ukowski for their help. This research has been completed before
arXiv:0710.3322 \cite{nowa} appeared. This work is part of EU 6FP programme QAP contract no. 015848.

\end{document}